\documentclass[]{mn2e}
\usepackage{graphicx}
\usepackage{epsfig}
\usepackage{amsmath}
\newcommand       \Angstrom     {\,{\rm \AA}}

\newcommand       \nm           {\,{\rm nm}}

\newcommand       \eV           {\,{\rm eV}}

\newcommand     \gtsim  {\lower.5ex\hbox{$\buildrel > \over \sim$}}
\newcommand     \ltsim  {\lower.5ex\hbox{$\buildrel < \over \sim$}}
\newcommand     \simgt  {\lower.5ex\hbox{$\buildrel > \over \sim$}}
\newcommand     \simlt  {\lower.5ex\hbox{$\buildrel < \over \sim$}}

\newcommand       \mum          {\,{\rm \mu m}}
\newcommand       \ppm          {\,{\rm ppm}}

\newcommand       \simali       {\sim\,}

\newcommand       \Cabs       {C_{\rm abs}}

\newcommand	  \NC           {N_{\rm C}}
\newcommand	  \NH           {N_{\rm H}}

\newcommand	  \CTOHCNT {\left[{\rm C/H}\right]_{\scriptsize\rm CNT}}
\title{On Carbon Nanotubes in the Interstellar Medium}
\author[Li et al.\ ]
       {Qi~Li$^{1,2}$\thanks{201531160001@mail.bnu.edu.cn},
        Aigen Li$^{2}$\thanks{lia@missouri.edu}, 
        B.W.~Jiang$^{1}$\thanks{bjiang@bnu.edu.cn}
        and 
        Tao Chen$^{3}$\thanks{taochen@kth.se}\\
        $^1$Department of Astronomy,
                Beijing Normal University,
                Beijing 100875, China\\
        $^2$Department of Physics and Astronomy,
             University of Missouri,
             Columbia, MO 65211, USA\\
          $^3$Department of Theoretical Chemistry and Biology, 
                 Royal Institute of Technology, 
                 Stockholm 10691, Sweden
             }
\begin{document}
\date{Accepted 2020 February 13. Received 2020 February 12; in original form 2020 January 15.}
\pagerange{\pageref{firstpage}--\pageref{lastpage}} \pubyear{2019}

\maketitle

\label{firstpage}
\begin{abstract}
Since their discovery in 1991, 
carbon nanotubes (CNTs) ---
a novel one-dimensional carbon allotrope ---
have attracted considerable interest worldwide 
because of their potential technological applications
such as electric and optical devices.
In the astrophysical context, 
CNTs may be present in the interstellar space
since many of the other allotropes of carbon
(e.g., amorphous carbon, fullerenes, nanodiamonds,
graphite, polycyclic aromatic hydrocarbons,
and possibly graphene as well)
are known to be widespread in the Universe,
as revealed by presolar grains in carbonaceous 
primitive meteorites and/or by their fingerprint 
spectral features in astronomical spectra. 
In addition, there are also experimental 
and theoretical pathways to the formation 
of CNTs in the interstellar medium (ISM).
In this work, we examine their possible presence 
in the ISM by comparing the observed interstellar 
extinction curve with the ultraviolet/optical
absorption spectra experimentally
obtained for single-walled CNTs of 
a wide range of diameters and chiralities.
Based on the absence in the interstellar extinction curve 
of the $\simali$4.5 and 5.25$\eV$ $\pi$-plasmon 
absorption bands which are pronounced
in the experimental spectra of CNTs,
we place an upper limit of $\simali$10$\ppm$ of C/H
(i.e., $\simali$4\% of the total interstellar C)
on the interstellar CNT abundance. 
\end{abstract}

\section{Introduction}\label{sec:intro}
As one of the most abundant elements in the Universe
only exceeded by hydrogen, helium and oxygen, 
carbon (C) is a major player in the evolutionary 
scheme of the Universe (Henning \& Salama 1998).
Since C atoms can form strong and stable single, 
double, and triple covalent bonds, they facilitate 
the generation of a large variety of allotropic forms,
including amorphous carbon, carbon chains, 
carbon nanotubes (CNTs), carbon onions, 
diamond, fullerenes (e.g, C$_{60}$, C$_{70}$), 
graphene, graphite, and polycyclic aromatic
hydrocarbons (PAHs).
While diamond and graphite have been known
for at least a few thousand years, the laboratory
discoveries of such low-dimensional 
carbon nanostructures as 
zero-dimensional (0D) fullerenes,
one-dimensional (1D) CNTs,
and two-dimensional (2D) graphene
were far more recent 
(see Dinadayalane \& Leszczynski 2010).

Thirty five years ago, C$_{60}$ and C$_{70}$ were 
serendipitously discovered by Kroto et al.\ (1985)
--- R.F.~Curl, H.W. Kroto and R.E. Smalley received
the 1996 Nobel Prize in Chemistry 
becasue of this spectacular discovery 
--- in the sooty residue 
experimentally generated by vaporising graphite
in a (hydrogen-lacking) helium atmosphere. 
Six years later, CNTs were discovered and synthesized
by Iijima (1991) as spin-off products of fullerenes,
using an arc-discharge evaporation method similar 
to that used by Kroto et al.\ (1985) 
for fullerene synthesis.
In 2004, the first isolation of a single graphene sheet
has been achieved by Andre Geim and Kostya Novoselov
who extracted single-atom-thick layers from bulk graphite
(see Novoselov et al.\ 2004). Because of this, they were
awarded the 2010 Nobel Prize in Physics.

The presence of many of these C allotropes 
in the interstellar medium (ISM) has been 
explicitly revealed or implicitly indicated
from presolar grains isolated from carbonaceous 
primitive meteorites (e.g., nanodiamonds, graphite;
see Lewis et al.\ 1987, Amari et al.\ 1990),
and/or from the observations of molecular 
and solid-state features in astronomical spectra 
and the realization that these features are 
linked to certain carbonaceous materials
(e.g., amorphous carbon, graphite, and PAHs;
see Stecher \& Donn 1965,
L\'eger \& Puget 1984,
Allamandola et al.\ 1985,
Pendleton \& Allamandola 2002,    
Qi et al.\ 2018).
While fullerenes, CNTs, and graphene are of
paramount importance in modern science and 
technology since these carbon nanomaterials 
provide exciting challenges and opportunites 
for physicists, chemists, biologists, engineers, 
and material scientists, 
in recent years, the astronomical community 
has also been passionate about their possible 
presence as well as the role they could have 
played in the interstellar and circumstellar space. 
It is worth noting that the initial synthesis 
of fullerenes was actually astrophysically 
motivated --- the experiments which resulted in
the discovery of fullerenes were originally aimed at 
understanding the mechanisms by which long-chain 
carbon molecules are formed in the interstellar space 
and circumstellar shells (Kroto et al.\ 1985).
A quarter of a century later, the detection of C$_{60}$ 
and C$_{70}$ and their ions in the interstellar and 
circumstellar space has been reported based on
their characteristic vibrational spectral bands
in the infrared (IR; Cami et al.\ 2010, 
Sellgren et al.\ 2010, Bern\'e et al.\ 2013,
Strelnikov et al.\ 2015). 
Webster (1992) suggested that fullerenes
and hydrogenated fullerenes (i.e., fulleranes)
could be responsible for an appreciable 
component of the optical and ultraviolet (UV)
interstellar extinction. 
It has also been proposed that single- and 
multi-shell fullerenes may be the carrier of
the still unidentified 2175$\Angstrom$ 
interstellar extinction feature
(e.g., see  Iglesias-Groth 2008, Li et al.\ 2008).
Cataldo (2002) measured 
the absorption spectra of fullerite 
(i.e., carbon soot containing fullerenes)
and C$_{60}$ fullerene photopolymer
in the wavelength range of 
$\simali$0.2--1.1$\mum$.
It was found that neither fullerite 
nor C$_{60}$ photopolymer
exhibits any strong absorption feature 
around $\simali$2175$\Angstrom$.
Instead, fullerite shows prominent
absorption bands at $\simali$2520--2670$\Angstrom$,
similar to that seen in some hydrogen-deficient 
and carbon-rich supergiant R CrB stars 
(e.g., R CrB, RY Sgr, V348 Sgr) which peaks 
around $\simali$2400--2500$\Angstrom$
(Hecht et al.\ 1984, Drilling et al.\ 1997).
Unlike fullerite, C$_{60}$ photopolymer
exhibits three absorption peaks at
$\simali$2710, 3890 and 5100$\Angstrom$
which are not seen in the ISM.

As the basic structural element of fullerenes,
graphene is intimately related to fullerenes
as demonstrated by Chuvilin et al.\ (2010) experimentally 
and by Bern\'e \& Tielens (2012) computationally
that C$_{60}$ could be formed from a graphene sheet.
%
However, whether graphene is present in
the interstellar and circumstellar space 
is less certain. 
Garc{\'{\i}}a-Hern{\'a}ndez et al.\ (2011, 2012)
detected a set of unusual IR emission features
in several planetary nebulae (PNe), 
both in the Milky Way and in the Magellanic Clouds,
and attributed them to planar C$_{24}$, a piece of 
graphene sheet.
Based on the absence of the 2755$\Angstrom$ 
absorption feature in the interstellar extinction curve 
characteristic of the $\pi$--$\pi^{\ast}$ electronic 
transition of graphene, Li et al.\ (2019) argued that 
in the ISM as much as $\simali$20$\ppm$ of C/H 
could be tied up in graphene 
(also see Chen et al.\ 2017).
In addition, Sarre (2019) attributed the widespread 
extended red emission 
(ERE; see Witt \& Vijh 2004)
--- a broad, featureless emission band 
at $\simali$5400--9500$\Angstrom$
--- to the photoluminescence of graphene
oxide nanoparticles.

Like fullerenes, graphene is also the basic 
structural element of CNTs:
CNTs can be envisioned as the result of rolling up 
a segment of a graphene sheet to form a seamless 
tubular structure of high aspect ratio. 
The possible presence of graphene in space 
therefore raises the exciting possibility that CNTs 
may also be present in the Universe. 
While the electronic UV and optical spectra 
and vibrational IR spectra of individual CNTs
depend on their diameters, lengths, and morphologies
(e.g., see Guo et al.\ 2004, Chen \& Li 2019, 
Weisman \& Kono 2019),
as will be elaborated below in \S\ref{sec:UVAbs},
the UV/optical absorption spectra of CNT mixtures of
different sizes and types remain essentially invariant
(Murakami et al.\ 2005).
As interstellar CNTs --- if present --- likely consist of
a mixture of many individual CNT species,
in this work we examine the possible presence of
CNTs in the ISM by confronting 
the experimentally-measured
UV/optical absorption spectra of CNT mixtures 
with the observed interstellar extinction curve
(see \S\ref{sec:extinction}).
The results are discussed in \S\ref{sec:discussion}
and summarized in \S\ref{sec:summary}.


\begin{figure*}
\centering
\includegraphics[width=0.5\textwidth]{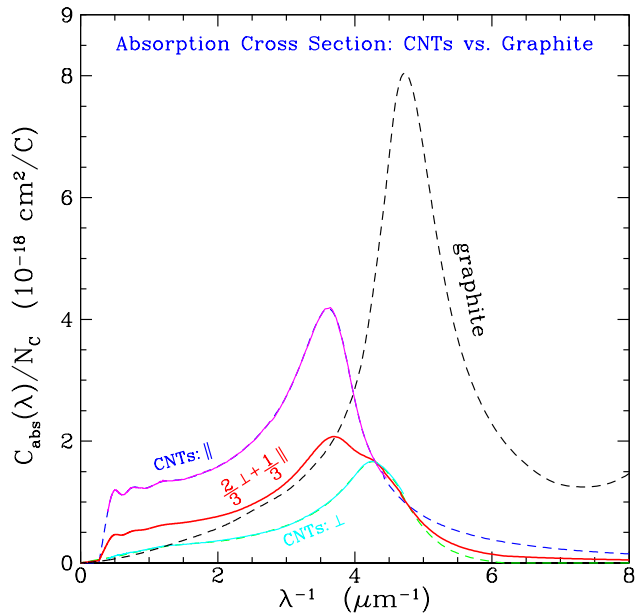}
\caption{\label{fig:Cabs}
         Experimental UV/optical absorption 
         cross sections (per C atom) of SWNTs 
         along the tube axis 
         [$\Cabs^{\parallel}(\lambda)/\NC$; solid magenta line]
         and perpendicular to the tube axis 
         [$\Cabs^{\perp}(\lambda)/\NC$; solid cyan line]
         obtained by Murakami et al.\ (2005)
         for vertically-aligned SWNTs 
         of various diameters and chiralities. 
         The mean absorption cross sections
         (solid red line) are obtained 
         with the ``1/3--2/3'' approximation.
         The measurements were made over 
         an energy range of $\simali$0.5--6$\eV$
         which corresponds to
         0.4\,$\simlt$\,$\lambda^{-1}$\,$\simlt$\,4.8$\mum^{-1}$.
         Extrapolations (dashed blue and green lines)
         are made for
         $\lambda^{-1}$\,$<$\,0.4$\mum^{-1}$ and
         $\lambda^{-1}$\,$>$\,4.8$\mum^{-1}$
         by fitting the experimental absorption spectra 
         at $\simali$0.5--6$\eV$ with a sum of
         multiple Drude functions.
         Also shown are the absorption cross sections 
         of nano graphite calculated from Mie theory
         using the dielectric function of Draine \& Lee (1984).
         }
\end{figure*}

\section{UV Absorption of Carbon Nanotubes}\label{sec:UVAbs}
Experimentally, CNTs exist in a varying number of  
concentric shells. Whereas the carbon-arc synthesis 
produces almost entirely multi-shell tubes 
on the carbon cathode, Iijima \& Ichihashi (1993)
found that abundant single-walled carbon nanotubes (SWNTs) 
with diameters of about one nanometer grow in the gas phase. 
As will be discussed in \S\ref{sec:discussion},
the possible pathways for the formation of CNTs 
in the ISM would result in SWNTs, 
more favorably than multi-walled nanotubes.
In this work, we will therefore focus on SWNTs.

SWNTs can be considered as a layer of  
graphene sheet rolled up into a cylinder.
The ends of each nanotube are closed 
by hemi-fullerene caps, each containing 
six five-membered rings.  
In the limit of large aspect 
(i.e., length-to-diameter) ratios,
however, the ends have negligible effects 
on nanotube electronic structure.
SWNTs are novel 1D materials made of 
an sp$^2$-bonded wall one atom thick
and constitute a rich family of structures,
with each SWNT structure uniquely defined 
by the chiral index, 
a  pair  of  integers ($n$,$m$),
which describes the length and orientation 
of the nanotube's circumference vector 
within the graphene sheet.
%
%
SWNTs are classified  into  three  types:
(i) armchair ($n$,$n$)  nanotubes,
(ii) zigzag ($n$,$0$)  nanotubes, and  
(iii) chiral ($n$,$m$) nanotubes 
with $n$\,$\neq$\,$m$.\footnote{%
  Armchair and zigzag SWNTs are identical 
  to their images and are therefore classified as
  achiral. Other SWNTs are chiral since they have 
  distinguishable mirror images (enantiomers) 
  of opposite handness.
  Chiral nanotubes can exhibit very large 
  (i.e., long) 1D unit cells compared to 
  achiral tubes of the same diameter. 
  }
%
%
%
%
%
Electrically, SWNTs can be either metallic 
or semiconducting, depending on their geometry, 
i.e., on the ($m$,$n$) integer-pair,
or on  the  way  of  the  rolling  up.
Armchair SWNTs are always metallic
and exhibit no energy bandgap, 
as are zigzag SWNTs with $m$\,=\,$3q$, 
where $q$ is an integer.\footnote{%
   Zigzag tubes can have 
   a small bandgap due to curvature effects.
   }
On a statistical basis, one-third of 
the nanotubes are metallic, 
and two-thirds are semiconducting. 

Each SWNT species, 
as labeled by a unique pair of integers ($n$,$m$),
has well-defined electronic 
and spectroscopic properties. 
Because of their distinct physical and chemical properties, 
different ($n$,$m$) structural species may be considered 
separate chemical substances. 
%
%
%
However, despite the importance of 
($n$,$m$)-specific absorption spectra,
reliable experimental determination of 
nanotube absorption cross section 
at the individual-tube level has been hampered 
by the difficulty of sorting as-grown mixtures 
into structurally pure fractions and by the challenge 
of determining absolute SWNT concentrations 
in suspensions that often also contain surfactants 
or polymer coatings (see Guo et al.\ 2004, 
Murakami \& Maruyama 2009, Weisman \& Kono 2019).
Whereas the interstellar extinction 
is the strongest in the UV, 
to the best of our knowledge, 
existing measurements of the absorption 
cross sections for individual SWNTs are limited 
to the visible and near-IR wavelength ranges
(e.g., see Liu et al.\ 2014, Streit et al.\ 2014,  
Sanchez et al.\ 2016, Yao et al.\ 2018).
Also,  interstellar SWNTs --- if present --- most likely
consist of a mixture of individual nanotubes
of different diameters and chiralities.
Therefore, in this work we will rely on
the absorption cross sections
measured by Murakami et al.\ (2005) 
for a vertically-aligned SWNT film
over the energy range of $\simali$0.5--6$\eV$
of astrophysical interest.
Other than that of Murakami et al.\ (2005), 
we were not aware of any other absorption 
cross section or dielectric function data 
available for ensemble nanotube samples
above $\simali$3$\eV$.

Murakami et al.\ (2005) grew the vertically-aligned 
SWNT film on an optically polished quartz substrate,
using the alcohol catalytic chemical vapor deposition 
(ACCVD) method. 
The film consists only of SWNTs that are sufficiently 
clean, i.e., contain virtually no amorphous carbon 
and no multi-walled carbon nanotubes, 
as confirmed by resonant Raman scattering  
and high-resolution 
transmission electron microscopy (HRTEM).
Direct HRTEM imaging measurements of 
more than 50 SWNTs revealed an average diameter 
of $\simali$2.0$\nm$ with a standard deviation of 
$\simali$0.4$\nm$. 
Most SWNTs in the film form bundles with 
a typical diameter of $\simali$15$\nm$,
while the thickness of the film is typically
$\simali$5--10$\mum$. 

SWNTs are optically anisotropic. 
Murakami et al.\ (2005) used 
the Shimadzu UV-3150 
spectrophotometer combined with 
a UV-vis-NIR polarizer to determine 
the polarization-dependent absorption spectra
of the vertically-aligned SWNT film
in the wavelength range of 
$\simali$200--2500$\nm$
which corresponds to an energy range of 
$\simali$0.5--6$\eV$.
As illustrated in Figure~\ref{fig:Cabs},
most noticeable in the absorption spectra 
are the broad, polarization-dependent peaks 
at $\simali$4.5 and $\simali$5.25$\eV$,
respectively arising from surface and bulk 
$\pi$-plasmon excitations.
The absorption feature at $\simali$4.5$\eV$ 
is seen for light polarized parallel to the SWNT axis, 
while the absorption feature at $\simali$5.3$\eV$ 
is seen for light polarized perpendicular to the axis. 
Murakami et al.\ (2005) found that 
these two peaks were observed at almost the same
positions regardless of the diameter and preparation 
method of SWNTs.
Due to intersubband absorptions,
in the low energy region ($<$\,3$\eV$)
three weak features at $\simali$0.63, 0.93
and 1.45$\eV$ are also seen in the absorption spectra
for light polarized parallel to the SWNT axis.\footnote{%
  These absorption features correspond to
   the first ($\simali$0.63$\eV$) 
   and second ($\simali$0.93$\eV$)
   subband gaps in semiconducting SWNTs 
   and the first gap ($\simali$1.45$\eV$) 
   in metallic SWNTs (Murakami et al.\ 2005).
   }

Let $\Cabs(\lambda)/\NC$ be the mean absorption
cross section per C atom of a SWNT of $\NC$ C atoms. 
Let $\Cabs^{\parallel}(\lambda)/\NC$ 
and $\Cabs^{\perp}(\lambda)/\NC$ respectively
be the absorption cross section per C atom 
of a SWNT of $\NC$ C atoms 
along or perpendicular to the tube axis. 
The ``1/3--2/3'' approximation 
(Draine 1988, 2016) leads to
\begin{equation}
\Cabs(\lambda)/\NC \approx 
\frac{1}{3}\,\Cabs^{\parallel}(\lambda)/\NC 
+ \frac{2}{3}\,\Cabs^{\perp}(\lambda)/\NC ~~.
\end{equation}
In Figure~\ref{fig:Cabs} we show
the polarization-dependent
absorption cross sections 
$\Cabs^{\parallel}(\lambda)/\NC$ 
and $\Cabs^{\perp}(\lambda)/\NC$ 
measured by Murakami et al.\ (2005) 
as well as the mean absorption
cross section $\Cabs(\lambda)/\NC$ 
determined from the ``1/3--2/3'' approximation. 
Also shown in Figure~\ref{fig:Cabs} is 
the extinction cross section of a graphite
nano particle of $\NC=40$ 
calculated from Mie theory 
(Bohren \& Huffman 1983)
using the dielectric functions of 
``astronomical graphite'' 
of Draine \& Lee (1984).
While the absorption cross sections
of SWNTs show two distinct peaks 
at $\simali$4.5$\eV$
(i.e., $\simali$2759$\Angstrom$ or 
$\simali$3.62$\mum^{-1}$)
and $\simali$5.25$\eV$
(i.e., $\simali$2365$\Angstrom$ or 
$\simali$4.23$\mum^{-1}$),
graphite exhibits a more pronounced 
absorption peak at $\simali$5.7$\eV$
(i.e., $\simali$2175$\Angstrom$ or 
$\simali$4.60$\mum^{-1}$).

\begin{figure*}
\centering
\includegraphics[width=.5\textwidth]{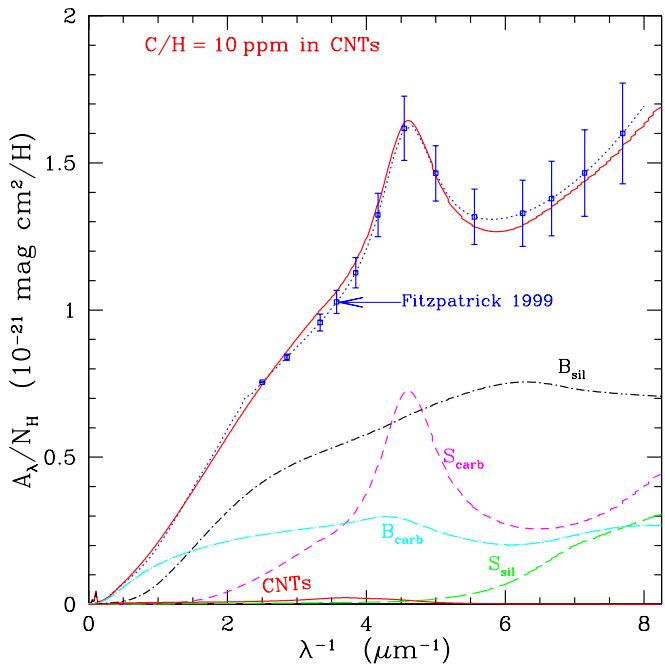}
\vspace{20mm}
\caption{\label{fig:extcurv}
        Comparison of the average Galactic interstellar extinction 
        curve (dotted line; Fitzpatrick 1999) 
        with the model extinction curve (solid red line)
        obtained by adding the contribution from CNTs
        with $\CTOHCNT=10\ppm$ (thin red line) 
        to the best-fit model of Weingartner \& Draine (2001).
        Also plotted are the contributions 
        (see Li \& Draine 2001) from 
        ``${\rm B_{sil}}$'' ($a$\,$\simgt$\,250$\Angstrom$ silicate);
        ``${\rm S_{sil}}$'' ($a$\,$<$\,250$\Angstrom$ silicate); 
        ``${\rm B_{carb}}$'' ($a$\,$\simgt$\,250$\Angstrom$ carbonaceous); 
        ``${\rm S_{carb}}$'' ($a$\,$<$\,250$\Angstrom$ carbonaceous, 
        including PAHs). 
        The vertical bars superposed on the extinction curve
        of Fitzpatrick (1999) represent the observational uncertainties.       
        }
\end{figure*}

\section{Extinction}\label{sec:extinction}
The interstellar extinction curve,
often expressed as the variation 
of the extinction $A_\lambda$ 
with the inverse wavelength $\lambda^{-1}$,
generally rises from the near-IR to the near-UV, 
with a broad absorption bump at 
$\lambda$\,$\approx$\,2175$\Angstrom$
or $\lambda^{-1}$$\approx$4.6$\mum^{-1}$, 
followed by a steep rise into the far-UV
at $\lambda^{-1}$$\approx$10$\mum^{-1}$,
the shortest wavelength at which the extinction
has commonly been measured 
(see Figure~\ref{fig:extcurv}).

The interstellar extinction curve contains 
important information about the interstellar 
dust size distribution and composition.
While SWNTs exhibit two prominent peaks 
at $\simali$4.5$\eV$ and $\simali$5.25$\eV$
in their absorption spectra 
(see Figure~\ref{fig:Cabs} and \S\ref{sec:UVAbs}),
the interstellar extinction curve 
is rather smooth in these energy ranges.
The {\it International Ultraviolet Explorer} (IUE)
obtained the extinction curves in the wavelength range
of $\simali$115--330$\nm$ along the lines
of sight toward hundreds of stars
and none of these extinction curves 
shows any structures resembling 
the 4.5 and 5.25$\eV$ peaks of SWNTs
(see Fitzpatrick \& Massa 2007).
Therefore, the nondetection of
the 4.5 and 5.25$\eV$ absorption features 
of SWNTs in the interstellar extinction curve
allows us to place an upper limit 
on the abundance of SWNTs in the ISM.

Weingartner \& Draine (2001) 
and Li \& Draine (2001) have developed
an interstellar grain model consisting of 
amorphous silicates, graphite, and PAHs.
This model closely reproduces 
both the Galactic interstellar extinction curve 
and the observed Galactic IR emission.
In Figure~\ref{fig:extcurv}
we show the extinction obtained 
by {\it adding} SWNTs to this model.
Let $\CTOHCNT$ be the amount of 
C (relative to H) tied up in SWNTs.
The extinction results from SWNTs
of a quantity of $\CTOHCNT$ is
\begin{equation}
\left(\frac{A_\lambda}{\NH}\right)_{\rm CNT}
= 1.086\,\left(\frac{C_{\rm abs}}{\NC}\right)_{\rm CNT}
\CTOHCNT ~~.
\end{equation}
As shown in Figure~\ref{fig:extcurv}, 
the maximum amount of SWNTs
allowable in the ISM is derived by requiring
the SWNT-added model extinction
not to exceed the observational uncertainties
of the Galatcic interstellar extinction curve 
(Fitzpatrick 1999).
In this way, we estimate the upper bound to 
be $\CTOHCNT\approx10\ppm$,
%
corresponding to $\simali$4\% of 
the total interstellar C
if we take the interstellar C abundance 
to be solar
(i.e., C/H\,$\approx$\,269$\pm$31$\ppm$,
Asplund et al.\ 2009).
The upper limit of $\simali$10$\ppm$
derived here for SWNTs is lower by 
a factor of $\simali$5 than PAHs 
(see Li \& Draine 2001).

\section{Discussion}\label{sec:discussion}
As fullerenes and CNTs are intimately related 
(e.g., nanotubes are often known as cylindrical fullerenes),
the ubiquitous detection of fullerenes in various
astrophysical environments (see Zhang \& Kwok 2013),
together with the remarkable stability of 
CNTs against intense radiation,
reinforces the idea that CNTs may also 
be widespread in the ISM.
Although the formation processes 
of fullerenes and CNTs
in space are still unclear to date,
they may both be related to PAHs
which are ubiquitously seen in 
a wide variety of astrophysical regions
as revealed by the distinct set of
IR emission features at 3.3, 6.2, 7.7,
8.6, 11.3 and 12.7$\mum$
(see Peeters 2014). 
%
%
Bern\'e \& Tielens (2012) proposed 
that fullerenes could be made in space
through graphene from the complete 
dehydrogenation of PAHs.  
Based on molecular dynamic simulations,
Chen et al.\ (2020) demonstrated that zig-zag SWNTs
could be formed from linear, catacondensed PAHs. 
Using pentacene (C$_{22}$H$_{14}$;
a linear, five-ring PAH molecule)
as a case study, Chen et al.\ (2020) 
found that vibrationally-excited 
linear PAHs could bend significantly 
without breaking the C--C bonds,
with a bending barrier easily surmountable
by absorbing a single stellar photon.
Following such a bending, a closed, 
stereospecific 3D nanostructure forms 
and could subsequently grow longer 
to zig-zag SWNTs 
via the conventional hydrogen-abstraction 
and acetylene-addition 
(HACA; Frenklach \& Feigelson 1989) process
which involves repetitive hydrogen losses 
from the 3D structure followed by the addition 
of one or two acetylene (C$_2$H$_2$) molecule(s).

Chen \& Li (2019) have also shown that armchair
 SWNTs could form in the ISM from benzene (C$_6$H$_6$) 
and phenyl (C$_6$H$_5$\text{\textbullet})\footnote{%
   Benzene has been detected in CRL~618, 
   a protoplanetary nebula (Cernicharo et al.\ 2001).
   More recently, McGuire et al.\ (2018) reported 
   the detection of benzonitrile ($c$-C$_6$H$_5$CN), 
   one of the simplest nitrogen-bearing aromatic molecules, 
   in the TMC-1 molecular cloud.
   }
through the HACA process. This formation pathway 
starts with a benzene combining with a phenyl radical 
to form a biphenyl radical (C$_{12}$H$_{10}$\text{\textbullet}) 
through hydrogen abstraction. Then, the biphenyl radical 
loses a hydrogen atom and reacts with another phenyl radical 
to form a triphenyl radical (C$_{18}$H$_{14}$\text{\textbullet}). 
The triphenyl radical, with the loss of a hydrogen atom,
 isomerizes to a closed 3D structure. 
Subsequently, the HACA process leads to the continual growth 
of the 3D structure to more complex armchair SWNTs.

Also, it has been experimentally shown that SWNTs could 
be catalytically formed on the surface of Fe nanoparticles 
of $\simali$1--10$\nm$ from any C-rich feedstock,
including CH$_4$, C$_2$H$_2$, benzene, and PAHs
(see Zhou et al.\ 2006 and references therein).
Nanotube length would depend on the local Fe/C ratio, 
and detailed chemical kinetics, and is expected to
show environmental variations.
In principle, a 1$\nm$-diameter SWNT 
could grow to lengths of hundreds of microns 
if gaseous reagents are continuously supplied 
(Hafner et al.\ 1998, Kong et al.\ 1998). 
In the ISM, shock processing could detach 
the formed SWNTs from the catalytic Fe nanoparticles 
and break long nanotubes into shorter ones.

For the catalytic formation of SWNTs 
on nanometer-sized Fe particles to be 
a feasible pathway in the ISM,
a prerequisite is the presence of
interstellar Fe nanoparticles.
In the Galactic ISM,
typically 90\% or more of the Fe is missing 
        from the gas phase (Jenkins 2009), 
        suggesting that Fe is the largest elemental
        contributor to the interstellar dust mass 
        after O and C and accounts for $\simali$25\% 
        of the dust mass in diffuse interstellar regions.
        However, as yet we know little about the nature 
        of the Fe-containing material.
        Silicate grains provide a possible reservoir 
        for Fe in the form of interstellar pyroxene
        (Mg$_x$Fe$_{1-x}$SiO$_3$) 
        or olivine (Mg$_{2x}$Fe$_{2-2x}$SiO$_4$).
However, the shape and strength of the 9.7\,$\mu$m 
        silicate feature in extinction suggest that 
        the silicate material is Mg-rich rather than 
        Fe-rich (Poteet et al.\ 2015) and therefore
        a substantial fraction ($\simali$70\%) of 
        the interstellar Fe could be in metallic iron
        or iron oxides (see Draine \& Hensley 2013). 
Thermodynamic computations have shown that 
Fe nanoparticles are stable against  sublimation in the interstellar
radiation field and can persist down to a radius 
of $\simali$4.5$\Angstrom$, and perhaps smaller
(see Hensley \& Draine 2017).
Therefore, interstellar Fe in the form of metallic Fe nanoparticles 
may indeed constitute a component of the interstellar dust.

In principle, the presence of CNTs 
in the ISM could be revealed through 
their C--C vibrational modes in the IR.
CNTs are actually more IR-active 
than graphene due to 
their cylindrical boundary condition.
However, as mentioned earlier,
the vibrational spectra of CNTs
depend on their diameters and chiralities
(e.g., see Chen \& Li 2019).
For SWNTs of diameters less than 2$\nm$, 
over 100 different structures are topologically possible.
Each different structural SWNT is a unique, ordered, 
molecular species having its own characteristic 
vibrational spectra. 
Therefore, a specific comparison of 
the vibrational spectra of individual SWNTs 
and the astronomical spectra 
is difficult because of the wide range of 
SWNT diameters and chiralities expected 
to be present in the ISM.
We hence call for experimental measurements
of the IR vibrational spectra of ensemble nanotube 
samples of a wide range of diameters, lengths,
and morphologies.
Kim et al.\ (2005) obtained the IR vibrational modes
of SWNTs based on the transmission spectra of
thin films of purified, freestanding SWNT bundles 
of typically $\simali$100 or more species with tube
diameters in the $\simali$1.2--1.6$\nm$ range 
grown by the electric arc method.
However, no information was provided on 
the strengths of the SWNT vibrational bands. 
This prevents us from any detailed modeling
of the IR emission of SWNTs 
in astronomical environments.

Electronic structure calculations have 
shown that SWNTs exhibit intense 
electronic transitions in the visible and near-IR, 
which vary systematically with length. 
Zhou et al.\ (2006) suggested that SWNTs 
might be responsible for some of 
the mysterious diffuse interstellar bands (DIBs).
The DIBs are a set of over 600 interstellar 
absorption spectral features first detected 
in 1919 (see Sarre 2006).
The identity of the species responsible for 
most of these bands remains as one of the most 
enigmatic mysteries in astrophysics.\footnote{%
   Campbell et al.\ (2015, 2016) and Walker et al.\ (2015)
   ascribed five DIBs at 9348.4, 9365.2,
   9427.8, 9577.0, and 9632.1$\Angstrom$
   to  C$_{60}^{+}$ (also see Foing \& Ehrenfreund 1994). 
   }
Both experimental and computational studies 
are needed to explore the electronic properties 
of individual SWNTs of various diameters and chiralities.
Also, due to the strong quantum-confinement 
($\simali$1\,nm) effect of charge carriers in SWNTs,
SWNTs luminesce efficicently 
over a broad wavelength region, 
ranging from the optical (Lin et al.\ 2005)
to the  near-IR (O'Connell et al.\ 2002).
It would be interesting to explore
whether SWNTs could account for
the unidentified ERE, an interstellar 
photoluminescence phenomenon (Witt \& Vijh 2004). 
Witt (2014) suggested that the ERE carriers 
and the carriers of DIBs may be connected.
Very recently, Lai et al.\ (2020) detected strong DIBs 
along the line of sight toward a background star
seen through the reflection nebula IC\,63, 
where the most intense ERE is detected. 
The detection of strong DIBs in association 
with ERE is consistent with the hypothesis 
of a common carrier for both DIBs and ERE
(Witt 2014).

\section{Summary}\label{sec:summary}
Motivated by the widespread of many allotropes 
of carbon (e.g., amorphous carbon, fullerenes, 
graphite, nanodiamonds, and PAHs) in the interstellar
and circumstellar medium, we have explored 
the possible presence of CNTs, 
another allotrope of carbon, 
in the diffuse ISM by comparing 
the experimentally-measured UV/optical 
absorption spectra of SWNT mixtures 
with the observed interstellar extinction curve. 
While the experimental absorption spectra 
of SWNTs exhibit two prominent peaks
at $\simali$4.5 and 5.25$\eV$
attributed to $\pi$-plasmon excitations,
the interstellar extinction curve is rather
smooth in this energy range.
Based on the absence of
the $\simali$4.5 and 5.25$\eV$ absorption
peaks in the interstellar extinction curve, 
we have placed an upper limit of 
$\simali$10$\ppm$ of C/H
(i.e., $\simali$4\% of the total interstellar C)
on the interstellar CNT abundance.
We call for further experimental measurements
and quantum-chemical computations 
of the UV/optical electronic and IR vibrational
transitions of CNTs of a wide range of 
sizes and chiralities.
%

\section*{Acknowledgements}
We thank Drs. X.H~Chen, B.T.~Draine,
G.Y.~Guo, J.~Kono, S.~Maruyama, Y.~Murakami, 
G.~Naik, D.S.~Wang, X.J.~Yang
and the referee
for very helpful discussions and suggestions.
This work is supported by NSFC through
Projects 11533002 and 11873041.
AL is supported in part by NSF AST-1816411.

\bsp
\label{lastpage}
\end{document}